\documentclass[11pt]{article}

\usepackage{jheppub}

\usepackage{epsf,epsfig,subfigure,graphicx}
\usepackage{color}

\newcommand{\dis}[1]{\begin{equation}\begin{split}#1\end{split}\end{equation}}
\newcommand{\be}{\begin{equation}}
\newcommand{\ee}{\end{equation}}
\def\bea{\begin{eqnarray}}
\def\eea{\end{eqnarray}}

\newcommand{\eq}[1]{Eq.~(\ref{#1})}

\newcommand{\bfrac}[2]{{\left(\frac{#1}{#2} \right)  }}\newcommand{\VEV}[1]{\langle #1 \rangle}

\newcommand\tev{\,{\rm TeV}}
\newcommand\gev{\,{\rm GeV}}

\newcommand\mdm{{M_{\rm DM}}}

\title{\Large\bf 
Decaying WIMP dark matter \\
for AMS-02 cosmic positron excess
}

\author[a,b]{Ki-Young Choi,}
\author[c]{Bumseok Kyae,} 
\author[a]{and~ Chang Sub Shin}

\affiliation[a]{Asia Pacific Center for Theoretical Physics, Pohang, Gyeongbuk 790-784, Korea}
\affiliation[b]{Department of Physics, POSTECH, Pohang, Gyeongbuk 790-784, Korea
}
\affiliation[c]{Department of Physics, Pusan National University, Busan 609-735, Korea}

\emailAdd{kiyoung.choi@apctp.org}
\emailAdd{bkyae@pusan.ac.kr}
\emailAdd{csshin@apctp.org}

\abstract{
For explaining the AMS-02 cosmic positron excess, which was recently reported, we consider a scenario of 
thermally produced and decaying dark matter (DM)
 into the standard model (SM) leptons with an extremely small decay rate, $\Gamma_{\rm DM}\sim 10^{-26}~{\rm sec.}^{-1}$. 
Since the needed DM mass is relatively heavy ($700~{\rm GeV}\lesssim m_{\rm DM}\lesssim 3000~{\rm GeV}$), 
we introduce another DM component apart from the lightest supersymmetric particle (``LSP'').
For its (meta-) stability and annihilation into other particles,  
the new DM should be accompanied with another $Z_2$ symmetry apart from the $R$-parity.
Sizable renormalizable couplings of the new DM with SM particles, which are necessary for its thermalization in the early universe, cannot destabilize the new DM because of the new $Z_2$ symmetry. 
Since the new DM was thermally produced, it can naturally explain the present energy density of the universe.
The new DM can decay into the SM leptons (and the LSP) only through non-renormalizable operators suppressed by a superheavy squared mass parameter after the new symmetry is broken around TeV scale.
We realize this scenario in a model of ``gauged vector-like leptons,'' which was proposed recently for the naturalness of the Higgs boson.  
}

\begin{document} 
\maketitle
\flushbottom




\section{Introduction} 
\label{sec:intro}


Recently, the Alpha Magnetic Spectrometer (AMS-02) collaboration has released their first observational result on the fraction of high energy cosmic positrons \cite{AMS02}.
Based on the $6.8\times 10^6$ events, AMS-02 collaboration has observed again the positron fraction excess $e^+/(e^++e^-)$ over the theoretical expectation in the energy range from 10 GeV to 350 GeV with unprecedented accuracy. 
It has confirmed the previous similar observation by  PAMELA \cite{PAMELA}, 
and seems to be consistent also with the cosmic $(e^++e^-)$ excess reported by Fermi-LAT \cite{FermiLAT}. 
Since the collaboration has not seen yet an anisotropy in the positron excess over the sky, 
the observation of AMS-02 seems to more support the idea that the positrons originate from dark matter (DM) in the halo rather than astrophysical sources. 

Thermally produced weakly interacting massive particles (WIMPs) have been long believed to be a leading candidate of DM, 
since they provide the correct order of magnitude of the cross section required for explaining the present energy density of the universe \cite{DMreview}. 
The lightest supersymmetric particle (LSP) in the minimal supersymmetric standard model (MSSM) has been regarded as one of excellent examples of the WIMP.
However, the recent observation on the positron excess is quite hard to be accommodated in the conventional framework of the thermally produced LSP DM scenario.  
Most of all, the leptonic annihilation channels of DM 
should be the dominant ones for the positron excess, 
but it is non-trivial because of the helicity suppression by the light leptons. 
It was noted that annihilation of Majorana fermions such as the LSP into $e_i^+$ and $e_j^-$, 
where $i,j$ are family indices, requires a too large boost factor,\footnote{In Ref.~\cite{HKK}, thus, co-annihilation between the LSP and another DM component was considered.} 
since the LSP should account for also the present relic density of the universe via such an annihilation process.
Moreover, if the DM mass is above 1 TeV and we accept the galactic profile of NFW or Einasto,  
the annihilation scenario would be disfavored \cite{TeVannihil,Cirelli:2008pk}.

On the other hand, in the DM decay scenario
a leptophilic decay of meta stable DM 
with a life time of $10^{26}~{\rm sec.}$ is assumed to be responsible for the positron excess.
%
%
The theoretical issues associated with the cosmic positron excess in the DM decay scenario would be 

{\bf (1)} how to obtain an extremely small coupling for the decay rate of $10^{-26}~{\rm sec.}^{-1}$, and   

{\bf (2)} how to naturally address the present relic density. 

\noindent 
Particularly, in the conventional supersymmetric (SUSY) DM models, {\bf (1)} is translated to the problem of how to get extremely small $R$-parity violation.   
It is known that such an extremely small decay rate can be achieved, 
if DM decay is dominated by a dimension 6 operator suppressed by a  squared mass parameter of order grand unified theory (GUT) scale \cite{Strumia}. 
In Ref.~\cite{Kyae}, it was noted that $R$-parity can be broken due to an electroweak (EW) scale vacuum expectation value (VEV) of a right-handed sneutrino ($\tilde{\nu}_1^c$), which interacts with the MSSM fields only through the superheavy SO(10) [or simply SU(3)$_c\times$SU(2)$_L\times$SU(2)$_R\times$U(1)$_{B-L}$ or U(1)$_{B-L}$] gauge bosons/gauginos.   
Hence, {\it thermally produced} bino-like LSP can decay to $e^+e^-\nu_1^c$ with the desired decay rate, 
if one right-handed neutrino ($\nu_1^c$) is light enough.\footnote{Even with {\it two} heavy right-handed neutrinos, the seesaw mechanism and the leptogenesis still work \cite{2seesaw}.}
As a consequence, the $U(1)_{B-L}$ breaking scale in SO(10) GUT can be determined with the data of the PAMELA's positron excess in this case.

Of course, if DM is the LSP, the relic density of the present universe can be naturally explained. 
However, if the DM mass is required to be relatively heavy, say $\gtrsim 1~{\rm TeV}$, all other super particles must be heavier than the LSP. 
Thus, a heavy LSP would spoil the status of SUSY as the solution of the gauge hierarchy problem. 
In this case, we need to introduce another heavier DM component, explaining the positron excess. 
If the interaction between the new heavier DM and the SM charged leptons  
is made extremely feeble for the meta-stability of DM, the relic density cannot be explained naturally. 
Due to the reason, non-thermal production of DM was broadly accepted for explaining the PAMELA and Fermi-LAT data. 
Unlike the WIMP scenario, however, the relic density of heavier DM is highly depends on the reheating temperature. 
Thus, carefully tuned reheating temperature should be necessarily assumed for explaining the present relic density.  

One way to avoid the problem ${\bf (2)}$ is to introduce two DM components $(\chi,X)$ \cite{Leptophilic}:  
the major DM component $\chi$ produced thermally is assumed to absolutely be stable, explaining the present relic density of the universe.     
On the other hand, the minor DM component produced non-thermally is meta stable, explaining the positron excess.   
As pointed out in Ref.~\cite{Leptophilic}, even quite small amount of leptophilic meta stable DM [$n_X/n_\chi>{\cal O}(10^{-10})$] can still explain the observed the positron excess, only if the decay rate is relatively larger.  

Another way to resolve ${\bf (2)}$ would be to introduce a new symmetry and assume that the new DM is the lightest particle among the fields charged under the new symmetry. 
Then the new DM can still remain stable, even if one turns on couplings between the new DM and other MSSM fields for thermalization of the new DM in the early universe.  
By controlling the breaking of the new symmetry, then,   
one can obtain the desired decay rate of the new DM.  
In this case, the new DM can decay to SM chiral ferimions (and the LSP) {\it only through non-renormalizable operators suppressed by a superheavy mass parameter} due to the new symmetry.
In this paper, we will focus on this possibility.

Recently, CMS and ATLAS have announced the discovery of the SM(-like) Higgs boson in the 125--126 GeV invariant mass range \cite{CMS,ATLAS}. 
In fact, 125 or 126 GeV is too heavy for the mass of the Higgs appearing in the MSSM. 
It is because such a heavy Higgs boson requires a stop  heavier than a few TeV in the MSSM, 
by which a fine-tuning of $10^{-3}-10^{-4}$ 
becomes unavoidable for explaining the $Z$ boson mass of 91 GeV. 
For the least tuning in the Higgs sector, thus, the stop should be as light as possible within the LHC bound ($\gtrsim 600~{\rm GeV}$), 
and the MSSM needs to be extended for explaining the observed Higgs mass. 
 
In Ref.~\cite{Vlepton}, the vector-like lepton doublets $\{L,L^c\}$, and singlets $\{N,N^c\}$ were introduced,
and their interaction with the Higgs in the superpotential  
\dis{
y_NLh_uN^c
} 
was considered. 
Like the top quark Yukawa coupling,  the Yukawa coupling $y_N$ could raise the radiative Higgs mass, if it is sizable.\footnote{If the vector-like charged lepton singlets, $\{E,E^c\}$ are also introduced, $h^0\rightarrow\gamma\gamma$ as well as the radiative Higgs mass can be enhanced.}  
Unlike the top quark coupling, however, $y_N$ of order unity would blow up below the GUT scale by the renormalization group (RG) effects, since the  gauge interactions associated with $\{L,N^c,h_u\}$ are too weak. 
Thus, an extra non-Abelian gauge symmetry $G$ [$=$SU(2)$_Z$] was also introduced, under which only the vector-like leptons $\{L,L^c;N,N^c\}$ are irreducible representations, while all the ordinary MSSM superfields remain neutral. 
In order to avoid the fine-tuning in the Higgs sector, 
their masses need to be lighter than 1 TeV. 
Fortunately, the LHC does not yet provide severe constraints on masses of extra vector-like leptons, 
if they eventually decay to the neutral components and SM chiral fermions. 
It is the reason why the vector-like leptons (rather than vector-like quarks) are seriously considered. 

We note here that the discrete $Z_2$ symmetry 
can always be embedded in $G$. 
It means that the lightest component among $\{L,L^c;N,N^c\}$ can be also a good DM candidate
apart from the conventional LSP, if it does not carry an electromagnetic charge.   
In the early universe, it could be in thermal equilibrium state with other MSSM fields through the Higgs and also SU(2) gauge bosons/gauginos. 
After it decoupled from the thermal bath, its relic density could support the energy density of the present universe together with the LSP. 
Its stability would be guaranteed by the $Z_2$ symmetry  embedded in $G$. 
However, $G$ should be eventually broken at low energies. As a consequence, the new DM component could decay, 
but only through a non-renormalizable operator suppressed by a superheavy mass parameter 
because of the gauge symmetry $G$, 
as will be seen later. 
%
%
%
%
%
%
%
So the decay should be extremely small in this case. 
Since the spontaneous breaking of $G$ leaves intact the $R$-parity, the LSP still remains absolutely stable.   

In this paper, we attempt to realize thermally produced and decaying DM without $R$-parity violation, explaining the observations by AMS-02, PAMELA, Fermi-LAT, etc., based on the theoretically well-motivated model.   
Since its mass is assumed to be around the EW scale or TeV for explaining the Higgs mass naturally, its thermal production could guarantee the desired quantity of the relic density of the present universe. 
Moreover, since the new DM doesn't have to be identified with the LSP, its relatively heavy mass needed for explaining them does not push up the mass spectrum of all the SUSY particles. 
The new DM component is theoretically well-motivated particle associated with the mechanism for the natural Higgs boson. 
Also the theoretically well-motivated new symmetry $G$ 
and its breaking make the new DM meta-stable, admitting a quite small decay rate of it.  
Although we will discuss the positron excess based on a specific model of Ref.~\cite{Vlepton}, 
it would be straightforward to generalize the mechanism discussed here.

The paper is organized as follows.  
In section \ref{sec:cosmicPositron}, we briefly describe the status of the observational results and related issues on cosmic positron excess.                
In section \ref{sec:Vleptons}, we review the model proposed in Ref.~\cite{Vlepton} for explaining the 126 GeV Higgs mass without a serious fine-tuning.
In section \ref{sec:DMprod}, we account for the present relic density in the framework of the model by estimation of thermally produced DM.
In section \ref{sec:DMdecay}, we discuss decay of the new DM, explaining the positron excess of AMS-02.
Section \ref{sec:conclusion} is a conclusion.


\section{Cosmic positrons from decaying dark matter} 
\label{sec:cosmicPositron}



The decaying DM has been studied in the literatures in order to explain the positron excess~\cite{Ibarra:2009dr,Cirelli:2008pk,Papucci:2009gd,Ackermann:2012rg,jp,Jin:2013nta,Kohri:2013sva}: leptophilic DM decay can produce the positron fraction of the PAMELA and AMS-02 as well as the electron-positron flux observed by Fermi-LAT without violating the constraint coming from anti-particle search.

For scalar DM case,  
it can account for the PAMELA and AMS-02 positron excess well, 
if its mass is around 1 TeV, lifetime is $5\times 10^{26}\sec.$, and the dominant decay channel is to $\mu^+\mu^-$ ~\cite{Cirelli:2008pk,Ibarra:2009dr,Ackermann:2012rg,twobody}.
However, 
if one attempt to accommodate also the $(e^+ + e^-)$ excess observed by the Fermi-LAT as well as the positron fraction within a common framework, 
the mass of DM should be increased up to 3 TeV~\cite{Ackermann:2012rg}.
For fermionic DM decaying into $l_i^+ l_j^-  \nu_k$, the DM mass needs to be above $500\gev$, and the lifetime longer than  $10^{26}\sec.$
to explain the data of PAMELA and AMS-02~\cite{Ibarra:2009dr,jp}. 
Both the Fermi $(e^+ + e^-)$ excess and PAMELA positron excess are nicely reproduced, if the fermionic DM mass is around 2.5 -- 3.5 TeV.\footnote{It was noted that in simple DM models, there is a tension between the AMS-02 positron excess and the Fermi electron-positron spectra, and there are studies on how to relax
 the tension \cite{tension}.}

Even decaying DM can also annihilate itself,  leaving gamma rays and anti-particles.
Such annihilation cross section of decaying DM would  be constrained mostly by the recent observation of gamma ray with the Fermi-LAT and anti-particles with PAMELA satellite.
Particularly, gamma-rays from DM annihilation could give stringent bounds on the DM annihilation cross section.  
Fortunately, however, e.g. if DM mass is about $1\tev$, it is  $1-2$ orders of magnitude higher than that needed for the thermal freeze-out relic density, depending on the annihilation modes~\cite{Ackermann:2011wa,Calore:2011bt,Ackermann:2012rg,Calore:2013yia}. Therefore, the annihilation effect of decaying DM in our scenario is safe from gamma-ray indirect detection.
However, cosmic rays created from DM decay could  interact with the interstellar medium and interstellar radiation field to produce photons through decay of pions, bremsstrahlung, and inverse Compton scattering. 
Thus, for instance, in the case of DM decaying into $\mu^+\mu^-$, the mass larger than 4 TeV is disfavored~\cite{Abdo:2010nz,Ackermann:2012rg}.

In some models of leptophilic three body decaying DM, the higher-order corrections such as radiative two-body decays  can produce photons and weak gauge bosons~\cite{Ciafaloni:2010ti,Garny:2010eg}. 
The resulting monochromatic photon lines or the hadronic particles could be constrained by observation.  
As will be seen later, however,  the final photons from lepton loop in our model is helicity-suppressed due to the small mass of the lepton, and so does not yield any remarkable effects. 

In the following sections, we will propose a specific model  to realize the above mentioned decay modes, 
and calculate the DM relic density and decay rate of DM in this model to show that it can successfully explain the positron excess.



\section{Gauged vector-like leptons} \label{sec:Vleptons}
 

For raising the radiative Higgs mass, we introduce the vector-like lepton doublets $\{L,L^c\}$, and neutral singlets $\{N,N^c; N_H,N_H^c\}$ with a gauge symmetry $G$. 
Their Yukawa coupling to the MSSM Higgs, and their mass terms are written as  
\dis{ \label{superPot}
W=y_NLh_uN^c + \mu_L LL^c + \mu_N NN^c + \mu_HN_HN_H^c , 
}
where $y_N$ is a dimensionless couplings. 
Such extra vector-like leptons, $\{L,L^c;N,N^c;N_H,N_H^c\}$ are assumed to be proper irreducible representations under $G$, 
whereas {\it all the ordinary MSSM superfields including the Higgs doublets remain neutral}.  
We will call them ``gauged vector-like leptons.''    
As a simple example, we consider the case of $G={\rm SU(2)}_Z$, and all the newly introduced vector-like leptons are the doublets under the SU(2)$_Z$. 
The mass parameters $\mu_L$, $\mu_N$, and $\mu_H$ in \eq{superPot} can be induced e.g. from the K${\rm\ddot{a}}$hler potential \cite{GM}, 
\dis{
K=\frac{X^\dagger}{M_P}
\left(\kappa_LLL^c+\kappa_NNN^c+\kappa_HN_HN_H^c\right) + {\rm h.c.} 
} 
Here $X$ denotes a SUSY breaking source: its $F$-component is assumed to develop a VEV of order $m_{3/2}M_P$. 
We suppose $|\mu_L|\gtrsim |\mu_N|$. 
The local and global quantum numbers for the relevant superfields are presented in Table \ref{tab:Qnumb1}. 
\begin{table}[!h]
\begin{center}
\begin{tabular}
{c|cccccc|c} 
{\rm Superfields}  &   ~$L$~   &
 ~$L^c$~  & ~$N$~  &  ~$N^c$~ & ~$N_H$ & ~$N_H^{c}$~ & ~$X$~ 
  \\
\hline 
SU(2)$_{Z}$  & ~${\bf 2}$ & ${\bf 2}$  & ~${\bf 2}$ & ${\bf 2}$
 & ${\bf 2}$ & ${\bf 2}$ & ~${\bf 1}$ 
 \\
 U(1)$_{R}$  & ~$1$ & ~$1$  & ~$1$ & ~$1$ 
 & $0$ & $2$  & ~$2$ 
 \\
 U(1)$_{PQ}$  & $-1$ & $-1$  & $-3$ & ~$1$ & $-1$ & $-1$
 & $-2$ 
\end{tabular}
\end{center}\caption{Matter fields charged under the {\it gauge} SU(2)$_{Z}$ and/or  
the global U(1)$_{\rm R}\times$U(1)$_{\rm PQ}$ symmetries.
The ordinary superfields of the MSSM are all {\it neutral} under SU(2)$_{Z}$. 
}\label{tab:Qnumb1}
\end{table}
Hence, there is no mixing between the extra vector-like leptons and the MSSM superfields except the $y_N$ term in \eq{superPot} at the renormalizable level. 
$\{N_H,N_H^c\}$ in \eq{superPot} play the role of the Higgs, breaking the SU(2)$_Z$ completely in the manner of the MSSM: 
the soft mass squareds of $\widetilde{N}_H$ and/or $\widetilde{N}_H^c$ can be negative at low energies through the RG evolutions, 
if they couple to other hidden matter with sizable Yukawa coupling constants, which we don't specify here. 
So $\widetilde{N}_H$ and/or $\widetilde{N}_H^c$ can develop non-zero VEVs of order TeV.

$\{L,N^c\}$ coupled to the Higgs $h_u$ make contributions to the radiative Higgs mass ($\equiv\Delta m_h^2$) as well as the renormalization of the soft mass squared of $h_u$ ($\equiv\Delta m_2^2$) \cite{Vlepton}:
\dis{
&~~ \Delta m_h^2|_{L,N^c}\approx N_V\frac{|y_N|^4}{4\pi^2}v_h^2{\rm sin}^4\beta
~{\rm log}\left(\frac{{M}^2+\widetilde{m}^2}{{M}^2}\right) ,
\\
&\Delta m_2^2|_{L,N^c}\approx N_V\frac{|y_N|^2}{8\pi^2}
\bigg[f_Q({M}^2+\widetilde{m}_l^2)-f_Q({M}^2)\bigg]_{Q=M_G} ,
}
where $N_V=2$ for $G={\rm SU(2)}_Z$ doublets, $v_h$ ($\equiv\sqrt{\VEV{h_u}^2+\VEV{h_d}^2}\approx 174~{\rm GeV}$) stands for the Higgs VEV with $\tan\beta=\VEV{h_u}/\VEV{h_d}$, and $f_Q(m^2)$ is defined as $f_Q(m^2)\equiv m^2\{{\rm log}(\frac{m^2}{Q^2})-1\}$. 
For simplicity, we set all the relevant soft mass squared to be the same as $\widetilde{m}^2$. 
$M^2$ denotes the mass squared of the fermionic component, $M^2\approx |\mu_L|^2+|y_N|^2v_h^2\sin^2\beta$.
%
%
%
The {\it quartic} power of $y_N$ in $\Delta m_h^2|_{\rm new}$ makes the radiative Higgs mass very efficiently raised, if $|y_N|$ is larger than unity.
Even for the stop mass squared of $\widetilde{m}_t^2\approx (600~{\rm GeV})^2$, thus,     
126 GeV Higgs mass can be easily explained,\footnote{The radiative correction by a heavy gluino ($\gtrsim 1~{\rm TeV}$) could make $\widetilde{m}_t^2$ too large at the EW scale. However, such an effect could be compensated by quite heavy other squarks ($\sim 10~{\rm TeV}$) via two loop effects \cite{U(1)'}. 
}  only if 
\dis{ \label{requirement} 
N_V|y_N|^4 
~{\rm log}\left(\frac{{M}^2+\widetilde{m}^2}{{M}^2}\right) ~ \approx ~  14.5,~5.4,~3.7,~2.9,~2.4    
}
for ${\rm tan}\beta=2,4,6,10,50$, respectively \cite{Vlepton}\footnote{In Ref.~\cite{Vlepton}, the analyses were performed with $\widetilde{m}_t^2\approx (500~{\rm GeV})^2$. 
Here we slightly change the numbers such that $\widetilde{m}_t^2\approx (600~{\rm GeV})^2$.} without ``$A$-term'' contribution. 
If $\widetilde{m}_t^2 > (600~{\rm GeV})^2$ [$\widetilde{m}_t^2 < (600~{\rm GeV})^2$], the left hand side of \eq{requirement} should be smaller [larger] than the right hand side for explaining the 126 GeV Higgs mass.
$\Delta m_2^2|_{L,N^c}$ is eventually associated with the fine-tuning issue, because it affects  determination of the $Z$ boson mass. 
In order to avoid a serious fine-tuning, $|\mu_L|^2$ and soft parameter $\widetilde{m}^2$ need to be as small as possible.
In Ref.~\cite{Vlepton}, it was assumed that $|\mu_L|^2$ and soft parameter $\widetilde{m}^2$ are smaller than $\widetilde{m}_t^2$ i.e. $(600~{\rm GeV})^2$. 
Actually, the introduction of the new vector-like leptons was possible, because the experimental bounds on the leptonic particles are not severe yet. 
In this paper, however, we take somewhat relaxed parameter space:
\dis{
|\mu_L|^2 ,~~\widetilde{m}^2 ~~\sim~~ (1~{\rm TeV})^2.
}    
   
Were it not for the new gauge symmetry $G$, $|y_N|$ of order unity at low energy would cause the Landau-pole problem below the GUT scale, i.e. 
$|y_N|$ blows up at a high energy scale by the RG effect, unless $|y_N|$ is quite smaller than unity at the EW scale, $|y_N| < 0.7$ (i.e. $|y_N|^4 < 0.24$).   
However, the extra gauge interactions from $G$ can efficiently protect the smallness of $|y_N|$ up to the GUT scale.

In order to maintain the gauge coupling unification, one can introduce also two copies of the vector-liked colored particles, $2\times \{D,D^c\}$. 
They are regarded as the singlets of SU(2)$_Z$, and so 
they compose $2\times \{{\bf 5},\overline{\bf 5}\}$ of SU(5) together with the SU(2)$_Z$ doublets $\{L,L^c\}$.  
It is assumed that $\{D,D^c\}$ don't couple to the Higgs, or they   
couple to the Higgs with relatively small Yukawa couplings such that the fine-tuning problem in the MSSM Higgs sector does not arise again. 
Alternatively, one can assume that their SUSY mass are large enough compared to their soft masses so that the radiative correction to the Higgs potential by them is suppressed. 
 
In Ref.~\cite{Vlepton}, one more pair of $\{{\bf 5},\overline{\bf 5}\}$ were introduced. 
But they don't play an important role except for affecting the RG evolutions.  
In this case, the {\it maximally allowed value} of $|y_N|$ at the EW scale is lifted up to 1.78 (i.e. $|y_N|^4 < 10$):
\dis{ \label{y_N}
|y_N| < 1.78 .
}   
Only if $|y_N|$ is smaller than $1.78$, thus, it does not blow up below the GUT scale.
We note that \eq{y_N} makes \eq{requirement} trivially satisfied. 
The most severe constraint comes from the EW precision test parametrized with $(\Delta S,\Delta T)$. 
For $|\mu_L|^2\approx \widetilde{m}^2$ ($\gg v_h^2$),
the lower mass bounds on $|\mu_L|$ turn out to be 803 GeV, 592 GeV, 517 GeV, 469 GeV, and 440 GeV for ${\rm tan}\beta=2$, 4, 6, 10, and 50, respectively.
In this parameter range, $\Delta S$ turns out to be $0.01\lesssim\Delta S\lesssim 0.02$ and $\Delta T\approx 0.12$, which corresponds to a range inside 1$\sigma$ band of $(\Delta S,\Delta T)$.  
Hence, $|\mu_L|^2$ and $\widetilde{m}^2$ can be smaller than $(600~{\rm GeV})^2$ for ${\rm tan}\beta > 4$. 

We note that in the superpotential \eq{superPot},  
the discrete $Z_2$ symmetry is embedded in the SU(2)$_{Z}$ gauge symmetry, 
and the odd parity of the $Z_2$ can be assigned to the extra vector-like leptons. 
It means the lightest component of the extra vector-like leptons is stable, and so can be a DM component. 
Particularly, if $|\mu_L|\gtrsim |\mu_N|$ as in Ref.~\cite{Vlepton}, 
the bosonic and fermionic fields of the neutral component of $\{N,N^c\}$ (and $\{L,L^c\}$) could be DM as well as the ordinary LSP in this model. 
Through the Yukawa interactions with the Higgs in \eq{superPot} and also the MSSM gauge interactions, they would be in thermal bath in the early universe.
We will seriously examine this possibility in the next section. 

However, SU(2)$_Z$ and so $Z_2$ should be broken at low energies by VEVs of $\{N_H,N_H^c\}$. 
Since the $Z_2$ is embedded in the SU(2)$_Z$ gauge symmetry, 
domain walls are not created. 
Note that SU(2)$_Z$ breaking does not leave monopoles.
But the conventional $R$-parity is not yet broken. 
Thus, the new DM could eventually decay to the SM chiral fermions (and the LSP), whereas the LSP remains absolutely stable. 
We should note here that they can decay only through non-renormalizable interactions due to the SU(2)$_Z$ symmetry as mentioned in Introduction. 
Accordingly, the new DM components can be thermally produced, but decay with quite small rate. 
We will discuss this mechanism in section \ref{sec:DMdecay}.  
With this scenario we will attempt to explain the recently reported AMS-02 cosmic positron excess.


\section{Thermal production of dark matter} \label{sec:DMprod}


%
The DM mass required for explaining the AMS-02 data based on DM decay scenario is relatively heavy ($\gtrsim 520-700~{\rm GeV}$). 
Hence we need to introduce other DM component(s) apart from the LSP.   
Otherwise, we should assume that all the SUSY particles are quite heavy, which would spoil the original motivation of low energy SUSY. 
As mentioned in section \ref{sec:Vleptons}, in our case the neutral components of $\{L,L^c;N,N^c\}$ could be new DM components. 
In addition to the ordinary LSP, thus, we have one or two more DM components, namely, 
the lightest bosonic and/or fermonic particles of the neutral components of $\{L,L^c; N,N^c\}$,   
if their lifetime is much longer than the age of our universe.

Although the AMS-02 cosmic positron excess will be explained with leptonic decay of meta stable DM, 
we first attempt to naturally account for the present relic density, based on the thermally produced WIMP scenario. 
For stability of the new DM, we need to introduce also a new symmetry embedding a new $Z_2$ apart from the $R$-parity. 
Otherwise, the new DM would immediately decay to lighter MSSM fields through a coupling introduced for its thermalization in the early universe. 
In our case, $G={\rm SU(2)}_Z$ ($\supset Z_2$) plays the role of such a new symmetry, even if $G$ should eventually be broken. 
Here we consider three kinds of DM components and calculate the freeze-out relic density of the DM components.

As seen in \eq{superPot}, the fermionic components of $\{L,L^c\}$ and $\{N,N^c\}$
have the Dirac masses, $\mu_L$ and $\mu_N$, respectively. 
The Higgs VEV mixes the neutral components of $\{L,N^c\}$. 
The squared masses for the heavier and lighter mass eigenstates are given by \cite{Vlepton} 
\begin{eqnarray} \label{evalues}
(M_{N_1})^2\approx |\mu_L|^2 + |y_N|^2|h_u|^2 ,
\quad (M_{N_2})^2\approx |\mu_N|^2-
\frac{|\mu_N|^2}{|\mu_L|^2}
|y_N|^2|h_u|^2, 
\end{eqnarray}
for $|\mu_L|\gtrsim |\mu_N|$. 
Since the Higgs VEV is much smaller than $\mu_L$ and $\mu_N$, from now on we will ignore the mixing effect:
the heavier (lighter) state is just the neutral component of $L^c$ ($N^c$) with 
the mass of $\mu_L$ ($\mu_N$).  

For the scalar components, there are additional soft term contributions such as
\dis{ \label{softLag}
&-{\cal L}_{\rm soft} = A_N \widetilde{L} h_u \widetilde{N}^c + B_L \widetilde{L} \tilde{L}^c + B_N \widetilde{N} \widetilde{N}^c + \textrm{h.c.} \\
&~~ + m_L^2 |\widetilde{L}|^2 + m_{L^c}^2 |\widetilde{L}^c|^2 + m_N^2|\widetilde{N}|^2+m_{N^c}^2|\widetilde{N}^c|^2.
} 
Thus, the lightest scalar mass is heavier than that of the fermionic one due to such a soft mass. 
Since the mass mixing is small enough,  
the lightest scalar as well as the lightest fermion dominantly come from the $N^c$ sector.

We should note that through the $y_N$ term in \eq{superPot} and the $A_N$ term in \eq{softLag}, 
the scalar component $\widetilde N^c$ can decay to 
the fermionic component $N^c$ and the higgsino $\tilde h_u$, if it is kinematically allowed. 
It is because $\widetilde N^c$ can be converted to $\widetilde L$ by the $A_N$ vertex after the Higgs gets the VEV, and $\widetilde L$ can further decay to $N^c$ and $\tilde h_u$ by the $y_N$ vertex.    
In this case only $N^c$ (and the LSP) contributes to DM. 
However, if the decay is kinematically forbidden,  there remains contributions from both $N^c$ and  $\widetilde{N}^c$. 
Hence, the present DM relic density is composed of the contributions from all the (meta-)stable fermionic and/or scalar DM components as well as the LSP.  Thus, the present relic density is
\dis{
\Omega_{\rm DM} h^2 \approx \sum_{i=N^c,\widetilde{N}^c,\tilde{\chi}^0} 0.1 \times \bfrac{2.57\times 10^{-9}}{\VEV{\sigma v}_i}.
}

\begin{figure}[t]
  \begin{center}
  \begin{tabular}{cc}
   \includegraphics[width=0.5\textwidth]{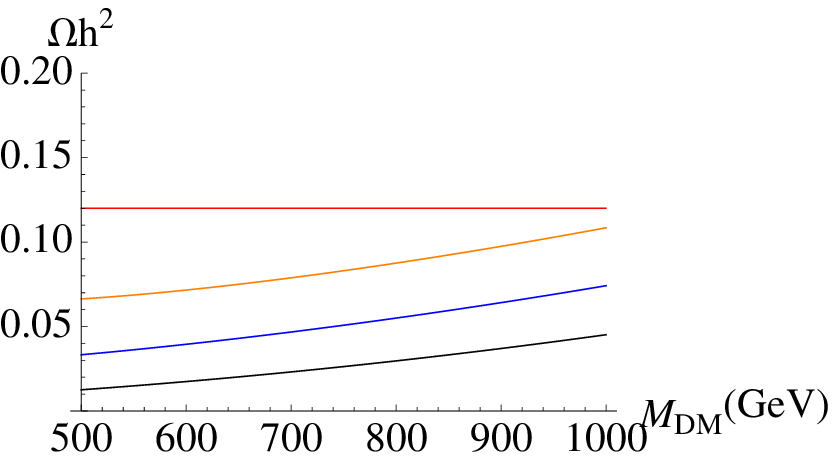}
   &
   \includegraphics[width=0.5\textwidth]{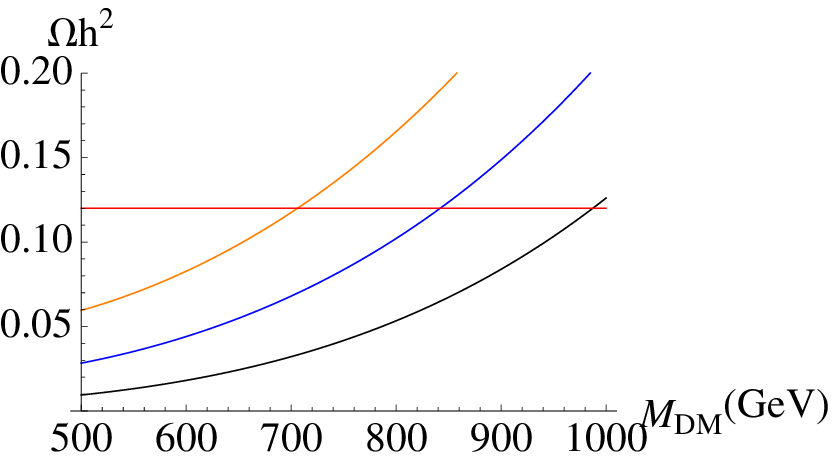}
  \end{tabular}
  \end{center}
\caption{ The freeze-out relic density of fermion (Left) and scalar (right) dark matter.  Here we used Yukawa coupling $y_N=1.5$ and different mass of $\mu_L-\mdm=50, 300,500 \gev$ (Black, Blue, Orange). For the scalar dark matter annihilation we used  higgsino mass $300\gev$. The horizon line (Red) is the present relic density of dark matter. }
\label{fig:Oh2}
\end{figure}

\underline{\bf Dirac fermion DM:}~
the lightest fermion $N^{c}$ annihilates into the Higgs $h_u$,  
via the $t$-channel process mediated by $L$, $N^{c}+\bar{N}^{c}\to h_u+h_u^*$.
The annihilation cross section is
\dis{
\VEV{\sigma v}_{N^c,\bar{N}^{c}}  \approx  \frac{y_N^4\mu_{N}^2}{64\pi(\mu_{N}^2+\mu_{L}^2)^2} . 
}
In figure~\ref{fig:Oh2}, we show the freeze-out relic density of the fermionic DM for different mass range.  For larger  mass of $N^c$, the annihilation cross section becomes smaller and the relic density increases. We assume that the neutralino LSP DM constitutes to the rest of dark matter needed 
for $\Omega_{\rm tot} h^2\approx 0.12$.  
%
%

\underline{\bf Complex scalar DM:}~
the complex scalars, $\widetilde{N}$ and $\widetilde{N}^c$  could also be stable. 
We suppose that $\widetilde{N}^c$ is the lightest scalar.
The $\widetilde{N}^c$ can annihilate into higgsino $\tilde{h}_u$ via the $t$-channel process mediated by $L$, 
$\widetilde{N}+\widetilde{N}\to \tilde{h}_u+\bar{\tilde{h}}_u$.
The annihilation cross section is given by 
\dis{
\langle\sigma v\rangle_{\widetilde{N}^c,\widetilde{N}^{c*}} \approx  \frac{y_N^4}{16\pi} \frac{ m_{\tilde{h}_u}^2}{(m^2_{\widetilde{N}^c}+\mu_L^2)^2}  + \frac{y_N^4}{12\pi} \frac{p^2}{(m^2_{\widetilde{N}^c}+\mu_L^2)^2} ~,
}
where $p$ ($\approx m_{\widetilde{N}^c} v_{DM}$) denotes the 3-momentum of $\widetilde{N}^c$.  
Using the relation $\VEV{v_{DM}^2} = 3 T/ (2m_{\widetilde N^c})$, we get $p^2 \approx \frac32 m_{\widetilde N^c}T$. 
At decoupling temperature,  
 $m_{\widetilde N^c}/T \approx 20 - 25$,  and so  $v_{DM}\sim 0.25 c$. 
The freeze-out relic density of the scalar DM is displayed in figure~\ref{fig:Oh2}.

\underline{\bf Ordinary neutralino LSP:}~
the total relic density of the universe should be around $\Omega_{\rm tot} h^2\approx 0.12$. 
We assume that apart from the contributions by $N^c$ and $\widetilde N^c$, the rest of the DM component needed for $\Omega_{\rm tot} h^2\approx 0.12$ is filled with the ordinary neutralino LSP DM.
This is naturally obtained for the light Higgsino LSP, since their annihilation cross section is large and the relic density is usually suppressed compared to the DM relic density~\cite{Edsjo:1997bg}.
%


\section{Decay of dark matter} 
\label{sec:DMdecay}


As explained in previous sections, the gauge symmetry $G$ [$={\rm SU(2)}_Z$] protects the stability of the new DM components, 
and makes their thermalization in the early universe possible. 
However, it is broken by VEVs of $\{N_H,N_H^c\}$.
On the contrary, {\it the $R$-prity is still conserved}.
After $G$ is broken, thus, the new DM components could decay to the SM chiral fermions and the LSP. 
%
%
%
Although $G$ is broken, its effect should still appear as $G$-invariant operators, 
where the VEVs of $\{N_H,N_H^c\}$ are involved. 
The VEVs of $\{N_H,N_H^c\}$ play the role of spurion fields. 
Thus, in the effective superpotential relevant to DM decay should include at least two newly introduced superfields for invariance under $G$: namely, one is DM $N^c$, and the other is one of $\{N_H,N_H^c\}$,   
which are irreducible representations of $G$. 
A scalar components of $\{N_H,N_H^c\}$ in the effective Lagrangian for DM decay should develop a VEV of order TeV, breaking of $G$ or $Z_2$.
For two or three body decay of DM, two or three MSSM superfields should couple to them. It implies that the effective superpotential for DM decay should be suppressed by a proper mass parameter. 
In order to obtain the extremely small decay rate, $\Gamma_{\rm DM}\sim 10^{-26}~{\rm sec.}^{-1}$, the decay amplitude needs to be suppressed by a squared mass parameter of order GUT, if the DM mass is around 1 TeV \cite{Strumia}.

For leptophilic decay of it, we will discuss two possibilities: 
an exotic quantum number for one of charged leptons could be responsible for a specific leptonic decay channel of the DM [``{Case (I)}''].
Heavy vector-like lepton pairs, which interact with visible leptons but are integrated out from low energy physics, can mediate small leptonic decay of the DM [``{Case (II)}''].

   
\underline{\bf Case (I):}~
we consider the case that the effective superpotential of DM decay is obtained after integrating out heavy singlet superfields $\{S,S^c\}$ from the following  superpotential:  
\dis{ \label{CaseI}
W_{\rm decay}^{(I)} = \lambda N^cN_HS + \frac{\kappa}{M_P} Z^2SS^c
+ \frac{\kappa_{ij}}{M_P}S^cl_il_je_1^c ,
}
where $\lambda$, $\kappa$, and $\kappa_{ij}$ are dimensionless couplings, and $M_P$ denotes the reduced Planck mass ($\approx 2.4\times 10^{18}~{\rm GeV}$). 
$Z$ is a spurion field carrying a global U(1)$_{PQ}$ charge. 
The VEV of $Z$ ($\sim 10^{12-13}~{\rm GeV}$) yields a mass parameter of $\{S,S^c\}$, $M_S\equiv \kappa\langle Z\rangle^2/M_P\sim 10^{6-8}~{\rm GeV}$,  
breaking the U(1)$_{PQ}$ symmetry completely. 
The global quantum numbers of the relevant superfields in \eq{CaseI} are presented in Table \ref{tab:Qnumb2}. 
\begin{table}[!h]
\begin{center}
\begin{tabular}
{c|ccc|cc|cc|cc} 
{\rm Superfields}  &  ~$Z$~ & ~$S$~ & ~$S^c$~ & ~$e^c_{i=1}$~
& ~$e_{i\neq 1}^c,d_i^c$~ & ~$\nu_i^c,u_i^c$ ~ & ~$l_i,q_i$~ & ~$h_{u}$~  &  ~$h_d$~  \\
\hline 
 U(1)$_{R}$ & ~$0$ & ~$1$ & ~$1$ & $-1$ & $-1$ & ~$1$ & ~$1$ & ~$0$ & ~$2$
 \\
 U(1)$_{PQ}$ & ~$1$ & ~$0$ & $-2$ & ~$0$ & ~$1$ & $-1$ & ~$1$ & ~$0$ & ~$-2$
\end{tabular}
\end{center}\caption{Global U(1)$_{R}\times$U(1)$_{PQ}$ charge assignment for some superfields neutral under SU(2)$_Z$ in Case I.
$Z$ is assumed to be a spurion superfield breaking U(1)$_{PQ}$. $i$ indicates the family index.  
}\label{tab:Qnumb2}
\end{table}

As seen in \eq{CaseI} and Table \ref{tab:Qnumb2}, 
U(1)$_{PQ}$ distinguishes $e_1^c$ from other mattr, which is a cause of the leptophilic decay of the DM.   
The charge assignment in Table \ref{tab:Qnumb2} permits not only the ordinary Yukawa couplings (except for the electron mass term) but also the following terms in the superpotential and K${\rm \ddot{a}}$hler potential:
\dis{
W\supset   y_i\frac{Z}{M_P}l_ih_de_1^c ~, ~~~{\rm and}~~~ K\supset \kappa_\mu\frac{X^\dagger}{M_P}h_uh_d + {\rm h.c.} ,
}
where $y_i$ and $\kappa_\mu$ denote dimensionless couplings.  
From the above terms, the electron mass term and the $\mu$ term can be generated after U(1)$_{PQ}$ symmetry and SUSY are broken, respectively. 
Since the VEV of $Z$ is of order $10^{12-13}~{\rm GeV}$, we can get the desired size of the Yukawa coupling for the electron mass ($\sim 10^{-5}-10^{-6}$). 
Note that any conventional $R$-parity violating terms are not allowed with this charge assignment. 
While the $R$-parity violating superpotential $l_il_je_k^c$, $q_il_jd_k^c$, and $d_i^cd_j^cu_k^c$ carry odd integer U(1)$_R$ charges ($\pm 1$), the assigned U(1)$_R$ charges of $\{N_H,N_H^c\}$ are only even integers ($0$,$2$). 
Accordingly, the $R$-parity breaking terms cannot generated perturbatively, even if U(1)$_R$ is broken by them.   

After integrating out $\{S, S^c\}$, the effective superpotential is obtained as\footnote{If the DM mass is around $260-270~{\rm GeV}$ and the life time is $\sim 10^{29}~{\rm sec.}$, one could try
a similar construction for DM decay
 to explain the Fermi's 130 GeV gamma ray line \cite{Weniger}, e.g.    
$W\supset N^cN_HW^\alpha W_\alpha/M_G^2$, 
where $M_G$ is a GUT scale mass parameter.  
Then the decay channels, $\widetilde N^c\to \gamma\gamma$ \cite{130gamma} and $\tilde\chi^0\to N^c\gamma$, etc. would be possible.    
However, we don't discuss this issue in this paper.} 
\dis{
W_{\rm eff}^{(I)}= - \frac{\lambda\kappa_{ij}}{M_S M_{\rm P}} N^c N_H l_il_je_1^c . 
} 
If the scalar component of $N^c$ is a DM component, 
it can decay eventually to three SM leptons and one neutralino by the above superpotential and the gaugino interaction.  
The positron energy spectrum in four body decay of DM
is expected to be relatively broad, and so a heavier DM would be required for coincidence with the observed spectrum. 
If $\widetilde{N}^c$ is quite heavier than the fermionic component $N^c$, ${m}_{\widetilde N^c}^2 > M_{N^c}^2+\mu^2$, however,  
$\widetilde{N}^c$ decays to $N^c$ and the higgsino (or eventually the neutralino LSP) via the Yukawa interaction \eq{superPot}. 
As mentioned in section \ref{sec:DMprod}, it is possible because $\widetilde N^c$ and the neutral scalar component of $L$ are mixed via the $A_N$ term in \eq{softLag}. 
For meta-stability of heavy $\widetilde{N}^c$, thus, $N^c$ should be also heavy.   
As mentioned above, however, heavy $\{\widetilde{N}^c,N^c\}$ are disfavored for the naturalness of the Higgs boson.  
In Case I, thus, we assume that their masses are relatively small, but satisfy 
${m}_{\widetilde N^c}^2 > M_{N^c}^2+\mu^2$ 
such that 
$\widetilde{N}^c$ decays to $N^c$ and $\tilde\chi^0$.

\begin{figure}[t]
  \begin{center}
  \begin{tabular}{c}
   \includegraphics[width=0.6\textwidth]{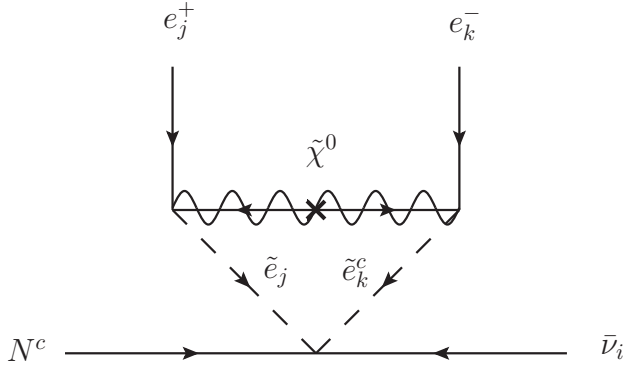}
  \end{tabular}
  \end{center}
\caption{Feynman diagram for decay of $N^c$ to $e^-e^+\bar\nu$ in Case I. $\tilde\chi^0$ is mostly wino-like.}
\label{fig:Nc_decay1}
\end{figure}

The dominant decay process of the fermionic component $N^c$ is the three body decay, $N^c \to e^- e^+ \bar\nu$ at one-loop level. 
See  figure~\ref{fig:Nc_decay1}.\footnote{One might think also a diagram that the $\tilde{e}_j$ and $\tilde{e}_k^c$ lines in figure~\ref{fig:Nc_decay1} 
directly merge into a Higgs line through the $A$-term vertex ($-{\cal L}_{\rm soft}\supset A^e_{jk}\tilde{l}_ih_d\tilde{e}_j^c+{\rm h.c.}$). 
However, such a diagram could be regarded as a suppressed one because of the smallness of the A-term coefficient or the Yukawa coupling, $A^e_{jk}\equiv y^e_{jk}m_{3/2}$. 
We neglect this diagram. 
} 
%
%
The relevant effective Lagrangian is given by 
\dis{
{\cal L}^{(I)}_{\rm eff}= \frac{\lambda\kappa_{ijk} \langle\widetilde{N}_H\rangle}{M_S M_{\rm P}} 
 N^c(\nu_i \tilde e_j - \nu_j \tilde e_i ) \tilde e^c_k + {\rm h.c.}  
}
%
In the limit of $m_{\tilde e^c}^2, m_{\tilde e}^2 \gg M_{N^c}^2, M_{\tilde\chi^2}^2$, thus, the initial spin averaged decay rate for $N^c\to \bar\nu_i e^+_j e^-_k$ is given by  
\dis{
&~ \Gamma(N^c\to \bar\nu_i e^+_j e^-_k)
\approx \frac{|\lambda\kappa_{ijk}|^2g_2^4}{24\pi (32\pi^2)^3} ~\frac{ M_{N^c}^5 \langle\widetilde{N}_H\rangle^2 M_{\tilde\chi^0}^2}{ M_S^2 M_{P}^2\widetilde m_{ e}^4}
~\left[{\rm log}\left(\frac{\widetilde m_{ e}^2 + \widetilde m_{ e^c}^2 }{\widetilde m_{e^c}^2}\right)\right]^2 
\\
&\approx 8.2\times 10^{-27}~{\rm sec.}^{-1}\times \left[\frac{|\lambda\kappa_{ijk}|^2}{10^{-2}}\right]
\left[\frac{\langle\widetilde{N}_H\rangle^2
M_{\tilde\chi^0}^2M_{N^c}^5/\widetilde m_{ e^c}^4}{(10^3~{\rm GeV})^5}\right]
\left[\frac{10^{8}~{\rm GeV}}{M_S}\right]^2
,
}
where we set the soft mass squareds of the left- and right-handed selectrons to be approximately the same, $\widetilde m_{e}^2\approx\widetilde m_{ e^c}^2$ for simplicity. 
With the above bench mark parameters, we achieve the desired decay rate of $10^{-26}~{\rm sec.}^{-1}$.  
For $M_{N^c}\approx 1~{\rm TeV}$ or smaller, we can successfully account for AMS-02 positron excess with less tuning in the Higgs sector.  
Assuming a heavier mass of DM, $M_{N^c}\approx 2-3~{\rm TeV}$, the $(e^++e^-)$ excess observed by Fermi-LAT can be also addressed, even if the fine-tuning of the Higgs boson becomes worse. 

As noted in \cite{Ciafaloni:2010ti,Garny:2010eg}, once the three body decay of $N^c\to\bar\nu_i e^+_j e^-_k$ is possible at one loop level, $N^c\to\bar\nu_i$ plus SM gauge bosons would be also possible at two loop. 
Such created photons or resulting hadrons would be  seriously constrained by observation. 
In our case, however, the chirality of the produced $e^+$, $e^-$ are opposite. 
Consequently, the photon and massive gauge boson productions in such a way should be helicity-suppressed due to a small charged lepton mass.



\underline{\bf Case (II):}~ we consider the following superpotential:
\dis{ \label{superPotB}
W_{\rm decay}^{(II)}=\lambda_i N^c l_i L_H^c + \kappa ZL_HL_H^c + \frac{\kappa_j}{M_P}N_HL_Hh_de_j^c , 
}
where $\lambda_i$, $\kappa$, and $\kappa_j$ are dimensionless couplings of order unity, and $Z$ is a spurion superfield developing a VEV of order $10^{10-12}~{\rm GeV}$. 
Here we introduced a new heavy vector-like pair of lepton doublets $\{L_H,L_H^c\}$, 
which can be responsible for a leptophilic decay of $\widetilde N^c$ or $N^c$. 
For the MSSM gauge coupling unification, one can introduce also vector-like quarks, $\{D_H,D_H^c\}$, 
but they can be forbidden to couple to the ordinary matter by the global symmetries in the lower dimensional operators.
Since $N^c$ and $\{L_H,L_H^c\}$ are charged under the non-Abelian gauge symmetry and $\{L_H,L_H^c\}$ are decoupled at the intermediate scale, even $\lambda_i$ (and $\kappa$) of order unity would not 
blowup below the GUT scale via the RG effects.      
The local and global quantum numbers of the relevant superfields in \eq{superPotB} are shown in Table \ref{tab:Qnumb3}.
\begin{table}[!h]
\begin{center}
\begin{tabular}
{c|ccccc} 
{\rm Superfields}  &  ~$N_H$~ & ~$N_H^c$~ & ~$L_H$~ & ~$L_H^c$~ & ~$e_{i}^c$~  \\
\hline 
SU(2)$_{Z}\times$SU(2)$_L$  & (${\bf 2},{\bf 1}$) & (${\bf 2},{\bf 1}$) & (${\bf 2},{\bf 2}$) & (${\bf 2},{\bf 2}$) & (${\bf 1},{\bf 1}$)
 \\
 U(1)$_{R}$ & $-1$ & ~$3$ & ~$2$ & ~$0$ & $-1$ 
 \\
 U(1)$_{PQ}$ & ~$0$ & $-2$ & ~$1$ & $-2$ & ~$1$ 
\end{tabular}
\end{center}\caption{Global U(1)$_{R}\times$U(1)$_{PQ}$ charge assignment for some matter fields in Case {II}.
The U(1)$_R$ charges of $\{N_H,N_H^c\}$ are modified from those in Table \ref{tab:Qnumb1}, but their $\mu$ term can still be induced from the K${\rm \ddot{a}}$hler potential after SUSY is broken by $X^\dagger$. 
The global charges of all the MSSM superfields except $e_i^c$ are the same as those in Table \ref{tab:Qnumb2}.
}\label{tab:Qnumb3}
\end{table}
Note that the U(1)$_R$ charges of $\{N_H,N_H^c\}$ in Case II are 
modified from those of Table \ref{tab:Qnumb2}. 
The global charges of $e_i^c$ follow those of $e_{i\neq 1}^c$ in Case I. 
The local and global quantum numbers of all other MSSM superfields are the same as those in Table \ref{tab:Qnumb1}.   
Again, the conventional $R$-parity violating couplings are disallowed. 
Still they cannot be induced perturbatively. 
 
\begin{figure}[t]
  \begin{center}
  \begin{tabular}{c}
   \includegraphics[width=0.9\textwidth]{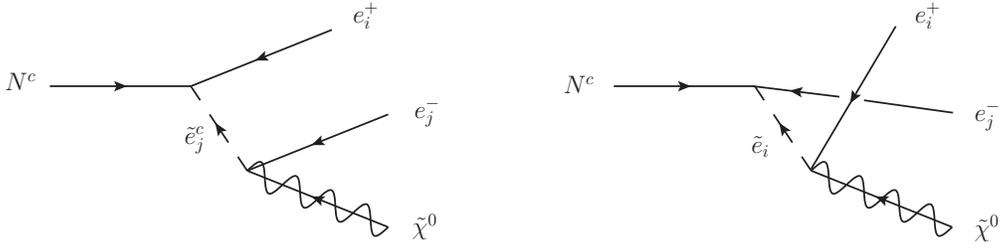}
  \end{tabular}
  \end{center}
\caption{Feynman diagrams for decay of $N^c$ to $e^-e^+\tilde\chi^0$ in Case II. $\tilde\chi^0$ is  wino or bino-like.}
\label{fig:Nc_decay2}
\end{figure}

After decoupling the heavy $\{L_H,L_H^c\}$, which are the bi-fundamental representation under SU(2)$_Z\times$SU(2)$_L$, 
we obtain the following effective Lagrangian for the decay of the bosonic and fermionic components of $N^c$:
\dis{
&\quad {\cal L}^{(IIB)}_{\rm eff}=  \frac{\lambda_i \kappa_{j}\langle\widetilde{N}_H\rangle v_h \cos\beta }{M_SM_{\rm P}} \widetilde N^c e_i e^c_j 
- \frac{\lambda_i \kappa_{j}\langle\widetilde{N}_H\rangle \sin\alpha}{\sqrt{2}M_S M_{\rm P}} \widetilde N^c h^0 e_i e^c_j 
+ {\rm h.c.} ,
\\
& {\cal L}^{(IIF)}_{\rm eff} = 
\frac{\lambda_i\kappa_{j} \langle\widetilde{N}_H\rangle v_h {\rm cos}\beta
}{M_SM_P}\left\{\frac{g_2}{\widetilde{m}_{l_i}^2}
(N^c e_j^c)(e_i\chi^0) +\frac{g_1}{\widetilde{m}_{e^c_j}^2}(N^c e_i)(e_j^c\chi^0)\right\}
+ {\rm h.c.} ,
}
where 
$g_{2,1}$ indicate the SM gauge couplings, and $h^0$ ($=\cos\alpha\, h_u^0 -\sin\alpha\, h_d^0 $) denotes the  Higgs boson.
Note that $M_S$ here is defined as $\kappa\langle Z\rangle\sim 10^{12}~{\rm GeV}$ unlike Case I.
$\widetilde{m}_{l_i}^2$ and $\widetilde{m}_{e^c_j}^2$
denote the soft mass squareds of slepton doublet and singlet, respectively.  
%
%
%
Unless ${m}_{\widetilde N^c}^2 > M_{N^c}^2+\mu^2$, the bosonic component $\widetilde{N}^c$ is meta-stable, and can be a DM component together with the fermionic component $N^c$. 
The bosonic component $\widetilde{N}^c$ can decay to two leptons, and also two leptons plus the Higgs.  
On the other hand, the dominant decay channels of the fermionic component are three body decays: 
when $\langle\widetilde{N}_H\rangle\neq 0$ and $\langle h_d\rangle\neq 0$ are involved, 
$N^c$ can decay, 
$N^c\to e^+\tilde e^-$, $\tilde e^+ e^-$, where the off-shell $\tilde e^-$ and $\tilde e^+$ are converted to $e^- \tilde\chi^0$ and $e^+ \tilde\chi^0$, respectively.   
See figure~\ref{fig:Nc_decay2}.  
%
%
%
%
If both $\widetilde{N}^c$ and $N^c$ are meta-stable,
thus, the decay rate of $\widetilde{N}^c$ is larger than that of $N^c$, because the mass of $\widetilde{N}^c$ and its kinematic factor in the decay rate are larger than those of $N^c$. 

Let us first estimate the rate of two body decay of $\widetilde{N}^c$, $\widetilde N^c\to \bar l_i \bar e^c_j$.  
In the limit of zero lepton mass, its decay rate is given by  
\dis{
&\qquad\qquad \Gamma(\widetilde N^c\to \bar l_i \bar e^c_j)= 
\frac{|\lambda_i\kappa_{j}|^2}{16\pi} ~\frac{ m_{\widetilde N^c} \langle\widetilde{N}_H\rangle^2 v_h^2 \cos^2\beta}{M_S^2 M_P^2} \\
&\approx 1.6\times 10^{-26}~{\rm sec.}^{-1}\times
\left[\frac{|\lambda_i\kappa_{j}|^2\cos^2\beta}{0.1}\right]\left[\frac{m_{\widetilde N^c}\langle\widetilde{N}_H\rangle^2}{(10^3~{\rm GeV})^3}\right]\left[\frac{10^{12}~{\rm GeV}}{M_S}\right]^2 .
}
Hence, one could successfully explain the AMS-02 positron excess 
with $\widetilde{N}^c$, if its mass is around 1 TeV.
Note that the decay rate just linearly depends on the DM mass $m_{\widetilde N^c}$ in this case. 
Accordingly, one can more easily increase the DM mass up to $2-3~{\rm TeV}$ such that the $(e^++e^-)$ excess of Fermi-LAT is also accommodated in this framework.  
As discussed in section \ref{sec:cosmicPositron}, however, the scalar DM decay of $\widetilde{N}^c\to e^+e^-$, $\tau^+\tau^-$ are disfavored because of the gamma ray constraint \cite{twobody}. 
In this case, thus, we should assume that  $\kappa_{22}$ is much larger than other components.  
$\mu^+\mu^-$ produced from $\widetilde{N}^c$ would subsequently decay to $e^+e^-$ and neutrinos, explaining the AMS-02 data. 

The three body decay rate of $\widetilde{N}^c$ ($\widetilde N^c\to \bar l_i \bar e^c_j h^0$) can be estimated as follows:  
\dis{
\Gamma(\widetilde N^c\to \bar l_i \bar e^c_j h^0)=
\frac{1}{96\pi^2}\left[\frac{m_{\widetilde N^c}^2}{v_h^2}\right]\left[\frac{\sin^2\alpha}{\cos^2\beta}\right]
I(x)\times \Gamma(\widetilde N^c\to \bar l_i \bar e^c_j) ,
}
where $x\equiv m_{h^0}/m_{\widetilde N^c}$, and $I(x)$ is the monotonically decreasing kinematic function,
\dis{
I(x) = (1-x^2)\left\{1+10 x^2+x^4+\frac{12 x^2(1+x^2)}{1-x^2}~{\rm log} x\right\} .
}
Note $I(0)=1$ and $I(1)=0$. 
The Higgs boson could further decay to hadrons, which would potentially be inconsistent with observations \cite{hadronic}. 
However, even if $m_{\widetilde N^c}=1~{\rm TeV}$ and so $m_{\widetilde N^c}^2/v_h^2 \approx 33$, the three-body kinematic suppression factor is
huge enough to be  
$\Gamma(\widetilde N^c\to \bar l_i \bar e^c_j h^0) \approx 0.026\, \Gamma(\widetilde N^c\to \bar l_i \bar e^c_j)$.
Consequently, the hadronic branching fraction of $m_{\widetilde N^c}$ decay is smaller than around 1\% of $\Gamma(\widetilde N^c\to \bar l_i \bar e^c_j)$
for $m_{\widetilde N^c}\sim 1~{\rm TeV}$.

On the other hand, if the scalar component of $N^c$ is heavy enough, ${m}_{\widetilde N^c}^2 > M_{N^c}^2+\mu^2$, it immediately decays to the fermionic component and the neutralino LSP. 
So only the fermionic component $N^c$ remains meta-stable. 
Its decay rate is estimated as 
\dis{
& \Gamma(N^c \to \tilde\chi^0 e^+_i e^-_j) =\frac{|\lambda \kappa_{ij}|^2}{16 \pi (384 \pi^2)}\left( \frac{ g_2^2M_{N^c}^5}{\widetilde m_{l_i}^4} + \frac{ g_1^2M_{N^c}^5}{\widetilde m_{e_j^c}^4}\right)
\left(\frac{ \langle\widetilde N_H\rangle^2 v_h^2 \cos^2\beta}{M_S^2 M_P^2}\right) 
I_2(x)
\\
&\approx 2.5\times 10^{-27}~{\rm sec.}^{-1}\times
\left[\frac{|\lambda \kappa_{ij}|^2\cos^2\beta}{10^{-2}}\right]\left[\frac{M_{N^c}^5\langle\widetilde N_H\rangle^2/\widetilde m_{e_j^c}^4}{(10^{3}~{\rm GeV})^3}\right]\left[\frac{10^{10}~{\rm GeV}}{M_S}\right]^2 I_2(x),
}
where we set the soft mass squareds of the sleptons to be approximately the same, $\widetilde m_{l_i}^2\approx \widetilde m_{e_j^c}^2$ for simplicity. 
The function of $x$ ($\equiv M_{\tilde\chi^0}/M_{N^c}$), $I_2(x)$ is given by 
\dis{
&\qquad\qquad I_2(x)= 
1- 8 x^2 -24 x^4 {\rm log} x +8 x^6 - x^8
\\
&+\frac{4 g_1 g_2  \widetilde m_{ l_i}^2 \widetilde m_{e^c_j}^2}{g_1^2 \widetilde m_{l_i}^4 + g_2^2\widetilde m_{e^c_j}^4}
\Big\{x + 12 (x^3+ x^5){\rm log} x + 9x^3 - 9 x^5 - x^7\Big\} ,
}
which approaches one for $x\ll 1$.    
For the above typical parameters, hence, we can obtain the decay rate needed for explaining the AMS-02 positron excess.  
Note that here we took $M_S\sim 10^{10}~{\rm GeV}$ unlike the case of the decay of $\widetilde N^c$.
Similar to Case I, $N^c \to \tilde\chi^0 \gamma$, etc. at one loop level, which is possible by contracting the $e^+$ and $e^-$ lines, are helicity-suppressed.

%
%


\section{Conclusion} 
\label{sec:conclusion}

In order to account for the AMS-02 cosmic positron excess and maintain low energy SUSY, it would be more desirable to introduce a new DM component apart from the conventional LSP. 
Since it is heavier than the LSP, an additional $Z_2$ symmetry should also be introduced in order to protect the stability of the new DM. 
This setup can be naturally embedded in the model of gauged vector-like leptons, which was proposed for the naturalness of the SUSY Higgs: 
the new DM component is embedded in the new gauged lepton $N^c$, and the additional $Z_2$ is in the SU(2)$_Z$ gauge symmetry in the model.

We have shown that 
the new DM thermalized with the MSSM fields in the early universe through its renormalizable couplings to  them
%
%
can naturally explain the present energy density of the universe. 
After the new $Z_2$ is broken around TeV scale, the new DM can decay into the SM leptons with the rate of $10^{-26}~{\rm sec.}^{-1}$ necessarily through non-renoramlizable operators suppressed by superheavy squared mass parameters, explaining the AMS-02 (and also Fermi-LAT) data. 
Since the $R$-parity still remain conserved, 
the conventional LSP DM is absolutely stable.
For leptophilic decay of DM, we proposed the two ways: 
one is to assign an exotic global charge to $e_1^c$,
and the other is to introduce a pair of heavy vector-like leptons interacting with the ordinary MSSM leptons.

\section*{Acknowledgments}

This research is supported by Basic Science Research Program through the 
National Research Foundation of Korea (NRF) funded by the Ministry of Education, Grant No. 2011-0011083 (K.-Y.C., C.S.S.) and 2013R1A1A2006904 (B.K.), and also in part by Korea Institute for Advanced Study (KIAS) grant funded by the Korea government 
(B.K.).
K.-Y.C. and C.S.S. acknowledge the Max Planck Society (MPG), the Korea Ministry of Education, Gyeongsangbuk-Do, and Pohang City for the support of the Independent Junior Research Group at the Asia Pacific Center for Theoretical Physics (APCTP).



\end{document}